\documentclass[epj,referee]{svjour}
\usepackage{amssymb}
\usepackage{graphics}
\usepackage{color}
\begin{document}
\title{Strangeons constitute bulk strong matter}
\subtitle{-- To test using GW~170817}
\author{Xiaoyu Lai,\inst{1,2} Enping Zhou\inst{3} \and Renxin Xu\inst{4,5}
}                     
%
%
\institute{School of Physics and Engineering, Hubei University of
Education, Wuhan 430205, China \and Research Center for Astronomy,
Hubei University of Education, Wuhan 430205, China \and
Max-Planck-Institut f\"ur Gravitationsphysik, Potsdam-Golm D-14471,
Germany \and School of Physics, Peking University, Beijing 100871,
China \and Kavli Institute of Astronomy and Astrophysics, Peking
University, Beijing 100871, China}
\date{Received: date / Revised version: date}
%
\abstract{The fundamental strong interaction determines the nature
of pulsar-like compact stars which are essentially in the form of
bulk strong matter.
From an observational point of view, it is proposed that bulk strong
matter could be composed of strangeons, i.e. quark-clusters with
there-light-flavor symmetry of quarks, and therefore pulsar-like
compact objects could actually be strangeon stars.
The equation of state (EOS) of strangeon stars is described in a Lennard-Jones model
for the purpose of constraining the EOS by
both the tidal deformability $\Lambda$ of GW~170817 and $M_{\rm TOV}$.
It is found that the allowed parameter space is quite large as most
of the Lennard-Jones EOS models satisfy the tidal deformability
constraint by GW170817.
The future GW detections for smaller values of $\Lambda$ and mass measurement
for larger values of $M_{\rm TOV}$ will help a better
constraint on the strangeon star model.
} 
\maketitle
\section{Introduction} \label{intro}
The strong matter we concentrate on in this paper refers to the
strongly interacting matter whose nature is determined by the
strong force~\cite{Xu2018}.
The most familiar form of strong matter to us is that of atomic
nuclei (with sizes $\sim$ fm).
In normal matter, nuclei are far way from each other, but the overall
properties of normal matter are controlled by the electromagnetic
force; however, this is not the whole story about the baryonic
matter in the Universe.

The bulk strong matter is macroscopic and the surface effect is
negligible\footnote{Similar to the case of strange quark matter, the
surface energy ($\propto R^2$) will become unimportant compared to
the bulk energy ($\propto R^3$) when the baryon number $A$ is large
enough. The ``bulk approximation'' is generally good for $A>10^2$
for strange quark matter~\cite{Madsen1994,Madsen1999}.}.
The lower limit of $A$ for bulk strange/strangeon matter, however,
is in fact not matter since we are concerning about the three-flavor
symmetric system.
The three-flavor symmetry would be restored in the strong matter
with size approximating to or even larger than the Compton
wavelength of electrons\cite{LX2017}, corresponding to baryon number
$A > 10^9$.
Therefore, the surface effect can be safely ignored for strong
matter with three-flavor symmetry, which is actually the bulk strong
matter.

Bulk strong matter could be produced by core-collapse supernovae of
evolved stars.
After core-collapsing of a massive star, the supernova-produced rump
is left behind, where normal nuclei are intensely compressed by
gravity to form the bulk strong matter, which could manifest in the
form of a pulsar-like compact object.

Nevertheless, the true nature of bulk strong matter is still
uncertain, which is essentially related to the ignorance about the
behavior of strong interaction at the low energy scales.
The neutron star and quark star are two models that have
attracted most attentions.
The former one originates from the concept of ``gigantic nucleus''
initiated by Landau~\cite{Landau1932}, and the latter compares
the whole star to a gigantic hadron composed of deconfined
quarks, based on the conjecture of Witten~\cite{Witten1984}.
From astrophysical points of view, however, it is proposed that
``strangeons'', which are formerly named as quark-clusters with
strangeness, could constitute bulk strong matter, and the
pulsar-like compact stars could actually be ``strangeon stars''
composed totally of strangeons.
The observational consequences of strangeon stars show that
different manifestations of pulsar-like compact stars could be
understood in the regime of strangeon stars (see the review
by~\cite{LX2017} and references therein).
More observational evidences to verify or disaffirm this proposal
are needed.

The gravitational wave event GW170817~\cite{ligo2017} and its
multiwavelength electromagnetic counterparts (e.g.,~\cite{Kasliwal2017})  open a new era in which
the nature of pulsar-like compact stars could be crucially tested.
The tidal deformability from the detection of gravitational waves
(GWs) from binary merger could put a clean and strong constraint on
the equation of state (EOS) of compact stars.
We have found that the tidal deformability of GW170817 and the
bolometric radiation could be understood if the signals come from
the merge of two strangeon stars in a binary~\cite{Lai2018RAA},
where the tidal deformability is derived from the EOS
in~\cite{LX2009}.
Further, it will be interesting and important to study
what the GW observation of tidal deformability means for EOS
 of strangeon stars and properties of strangeon matter, by the constraints on model parameters.

This paper is organized as follows: In \S \ref{sec:2} we briefly
introduce the concept of strangons constituting the bulk strong
matter, and the EOS of strangeon stars in a Lennard-Jones model.
In \S \ref{sec:3} we derive the dependence of tidal deformability of
merging strangeon stars
on the parameters in the Lennard-Jones model~\cite{LX2009},
and the constraint by GW170817.
Conclusions and discussions are made in \S \ref{sec:4}.

\section{The bulk strong matter} \label{sec:2}

The dense matter inside pulsar-like compact stars is strong matter
because the average density should be supra-nuclear
density (a few nuclear saturation densities) due to gravity.
The Fermi energy of electrons are significant in compressed baryonic
matter, and it is very essential to cancel the  energetic electrons by weak
interaction in order to make a lower energy state.
There are two ways to eliminate electrons.
The conventional way is via $e^-+p\rightarrow n+\nu_e$ as suggested in popular
neutron star models (i.e., {\it neutronization}).
On the other hand, a 3-flavor symmetry of quark could be restored in strong matter,
since the energy scale ($>\sim 400$ MeV) is
much larger than the mass difference between $s$ and $u/d$
quarks.
Consequently, another possible way to eliminate electrons could be
through the so-called {\it strangenization},
which is related to the flavor symmetry of strong-interaction matter.
Strangenization has both the advantages of minimizing the electron's
contribution of kinetic energy and maximizing the quark-flavor number.
%


\subsection{Strangeon and strangeon star} \label{sec:2.1}

If dense matter changes from a hadronic phase to a deconfined phase
as baryon density increases, the strong matter in compact stars
could be strange quark matter.
As stated by Witten~\cite{Witten1984}, if strange quark matter in
bulk may constitute the true ground state of strong matter rather
than $^{56}$Fe, then compact stars could actually be strange quark
stars instead of neutron stars.
However, the problem is: can the density of realistic
compact stars be high/low enough for quarks to become
deconfined/confined?

The state of compressed baryonic matter is essentially relevant to the
non-perturbative chromodynamics (QCD) problem,
and at the realistic density of
compact stars the quarks should neither be free nor weakly
coupled.
Although some efforts have been made to
understand the state of pulsar-like compact stars in the framework
of conventional quark stars, including the MIT bag model with almost
free quarks~\cite{Alcock1986} and the color-superconductivity state
model~\cite{Alford2008}, realistic stellar densities cannot be high
enough to justify the use of perturbative QCD which most of compact
star models rely on.

The bulk strong matter whose density is higher than the nuclear
matter density is proposed to be strangeon matter. This can be
understood in two approaches. In the approach from free quark state
(a top-down scenario), the strong coupling between quarks may
naturally render quarks grouped in
quark-clusters~\cite{Xu2010,LX2017}; and in the approach from
hadronic state (a bottom-up scenario), it is the {\it
strangeonization} to convert nucleons into strangeons, instead of
the {\it neutronization} that convert protons to neutrons, during
compressing normal baryonic matter of core-collapse supernova.
Each quark-cluster is composed of several quarks condensating in
position space rather than in momentum space.
Quark-cluster with three-light-flavor symmetry is renamed
``strangeon'', being coined by combining ``strange nucleon'' for the
sake of simplicity.

Bulk strangeon matter may constitutes the true ground state of
strong-interacting matter rather than nuclear matter
~\cite{LGX2013}.
%
%
This proposal could be regarded as a {\it general Witten's
conjecture}: bulk strange matter could be absolutely stable, in
which quarks are either free (for strange quark matter) or localized
(for strangeon matter).
Due to both the strong coupling between quarks and
the weak interaction,
the pulsar-like compact stars could be actually
strangeon stars which are totally composed of strangeons.
A strangeon star can then be thought as a 3-flavored gigantic
nucleus, and strangeons are its constituent as an analogy of
nucleons which are the constituent of a normal (micro) nucleus.

Different manifestations of pulsar-like compact objects
have been discussed previously
(see a review by~\cite{LX2017} and references therein)
in the strangeon star model.
Strangeon stars could help us to naturally understand the
observations of pulsar-like compact stars, both their surface and
global properties, for example, the drifting and bi-drifting sub-pulses~\cite{Xu1999},
the clean fireball for core-collapse supernovae and cosmic gamma-ray bursts (GRBs)~\cite{CYX2007},
the neutrino burst during SN 1987A~\cite{Yuan2017},
the spectra of XDINSs from optical to X-ray bands~\cite{Wang2017},
the high-mass pulsars~\cite{LX2009,LGX2013,GLX2014},
the radiation of anomalous X-ray pulsars (AXPs) and soft gamma-ray repeaters
(SGRs)~\cite{XTY2006,Tong2016},
and the glitch behavior of pulsars~\cite{Lai2018MN}.
It is also worth noting that, although the the EOS is very stiff,
the causality condition is still satisfied for strangeon matter~\cite{Lu2018}.

Moreover, the recently observed gravitational waves GW170817~\cite{ligo2017} as well as the electromagnetic radiation (e.g.,~\cite{Kasliwal2017}) could be understood if the signals come from the merge of two strangeon stars in a binary~\cite{Lai2018RAA}.
The tidal deformability is derived in the Lennard-Jones model~\cite{LX2009}, where the interaction between strangeons are assumed to be similar to that between molecules of inert gas.

\subsection{EOS of strangeon stars in Lennard-Jones model} \label{sec:2.2}

As stated above, pulsar-like compact stars could actually be
strangeon stars, where strangeons form due to both the strong and
weak interactions and become the dominant components inside those
stars.
Similar to a nucleon, a strangeon is composed of constituent quarks,
but there are two differences: the strangeon is of 3-flavored, and
the number of constituent quarks could be large than three.
Although we have proposed that $H$-dibaryons (with structure
$uuddss$) could be a possible kind of strangeons~\cite{LGX2013},
what could be the realistic strangeons inside compact stars is
uncertain due to the difficulties in QCD calculations.

As shown by Wilczek~\cite{Wilczek2007}, the interaction between
nucleons are characterized by the long-range attraction and
short-range repulsion.
Although the Lennerd-Jones potential originally describe the
interaction between inert gas molecules, it also have the character
of long-range attraction and short-range repulsion.
In this paper, we use a more general and phenomenological model, the
Lennard-Jones model~\cite{LX2009}, to describe the EOS of strangeon
stars and to find out the constraints from the tidal deformability
of GW170817.

In the Lennard-Jones model, the interaction between strangeons are
assumed to be similar to that between molecules of inert gas, since
strangeons are colorless as in the case of chargeless
atoms\footnote{It is worth noting that nucleon (2-flavored) and
strangeon (3-flavored) are two kinds of the colorless strong units
as atom of chargeless electric unit, and it would not surprising
that both nucleon/strangeon and atom could share a common nature of
6-12 potential.}.
The dependence of the potential $u$ on the distance between
strangeons $r$ is
\begin{equation}
u(r)=4U_0[(\frac{r_0}{r})^{12}-(\frac{r_0}{r})^6],
\end{equation}
where $U_0$ is the depth of the potential and $r_0$ can be
considered as the order of interaction range.
This form of potential has
the property of short-distance repulsion and long-distance
attraction, like the interaction between nucleons which stems
from the residual chromo-interaction.
By the approximation that only the two nearby strangeons have interaction
to each other, the EOS of strangeon stars can be derived under the above potential,
and the details are given in~\cite{LX2009}.

At the late stage of merging strangeon stars, the temperature should be $>\sim 10$ MeV
due to the tidal heating.
As a result, although an isolate strangeon star could be in the solid state~\cite{Dai2011} at low temperature,
the strangeon stars in a binary just before merger could be in the fluid state.
Consequently, to calculate the tidal deformability in the next section,
we neglect the contribution from the lattice vibrations~\cite{LX2009} to the EOS.

The energy density is then
\begin{equation}
\epsilon=2U_0(A_{12}r_0^{12}n^5-A_6r_0^6n^3)+nmc^2,
\end{equation}
and the pressure is
\begin{equation}
P=4U_0(2A_{12}r_0^{12}n^5-A_6r_0^6n^3),
\end{equation}
where $n$ is the number density of strangeons, $m$ is the mass of each strangeon.
If the number of quarks inside each strangeon is $N_q$, then we could approximate
that $m \simeq N_q\cdot 300$ MeV, where $N_q=18$ in the following calculations.
In addition, $A_{12}$ and $A_6$ are coefficients, relating to the
micro-structure of strangeon matter.

At the late stage of coalescence of binary strangeon stars, the
stars would melt by the tidal heating, but we still adopt
$A_{12}=6.2$ and $A_6=8.4$ for simplicity as in the case of the
simple-cubic structure, since other choices would not bring
significant changes.
The Lennard-Jones model reflects an important feature of strangeon
matter, i.e. the long-range attraction and short-range repulsion
between strangeons, no matter the strangeon matter is in the solid
or liquid state.
The short-range repulsion plays the crucial role in stiffening the
EOS and raising the maximum mass.
The form of EOS will not change significantly when the matter
changes from the solid to liquid state, although the specific values
of $A_{12}$ and $A_6$ should change since they are determined by the
micro-structure.

Moreover, although the values of $A_{12}$ and $A_6$ will also affect
the tidal deformability of strangeon stars, the quantitative results
remain unchanged when we choose some different values of $A_{12}$
and $A_6$ (but not differ by the order of magnitude) for liquid
stars.
Other choices of $A_{12}$ and $A_6$ would not change the result that
the tidal deformability of (liquid) strangeon stars are very
different from that of neutron stars, and the allowed parameter
space is quite large for the Lennard-Jones EOS models to satisfy the
tidal deformability constraint by GW170817.

It is also worth mentioning that, as discussed in \S\ref{sec:4},
there would be a sudden increase in the tidal deformability
resulting from the phase transition.
Qualitatively, this change is due to the differences in breaking
strain and shear modulus between solid and liquid states, regardless
of what specific values of parameters we choose.

Besides the different compositions, there is another difference
between neutron stars and strangeon stars, i.e. the surface
densities, which also affect the global structure of the stars.
Neutron stars are gravity-bound, while strangeon stars are
self-bound (similar to strange quark stars, and the self-bound
nature of strangeon stars is helpful to understand the drifting
sub-pulses).
Consequently, neutron stars have negligible surface density, while
strangeon stars have the surface density that higher than nuclear
matter density.
Although it seems that the hadronic matter can also be described by
Lennard-Jones model and have the corresponding form of EOS, the
global structures of neutron stars and strangeon stars are still
different.

The parameters $U_0$ and $r_0$ included in the EOS characterize the inter-strangeon potential.
The potential in Wilczek's paper~\cite{Wilczek2007} has a well with
the depth about 100 MeV, so in the calculation of \S\ref{sec:3}, we
choose the range of $U_0$ to be from 20 MeV to 100 MeV.
The surface number density of strangeons $n_{\rm s}$ determines $r_0$ by the fact that the pressure vanishes at the surface.
When translating $n_{\rm s}$ into the rest-mass density of strangeon
matter on the surface $\rho_{\rm s}=mn_{\rm s}$, we can constrain
$U_0$ and $\rho_{\rm s}$ from the EOS-dependent observable
properties.
The constraints by the mass-radius curves are discussed
in~\cite{LX2009}, and the TOV maximum mass could be higher than
$3M_\odot$.

The majority of pulsar-like compact stars are produced in
core-collapse supernovae, which usually have massed around $\sim 1.5
M_\odot$. More massive ones with masses approach or beyond
$2M_\odot$ are produced in binary star mergers and binary systems
with high accretion rates (e.g. some Ultra-Luminous X-ray sources),
so the birth rate is much lower. Therefore, although the theoretical
TOV maximum mass of pulsar-like compacts in strangeon star model
could above $3M_\odot$, the most detected ones are below $2M_\odot$.
In the era of multi-messenger astronomy, gravitational wave events
from binary star mergers, like GW~170817, could give better
constraints of the maximum mass and test various models.

In the next section we will show the constraints by both the maximum
mass of a static compact star ($M_{\rm TOV}$) and the tidal
deformability of GW~170817.

\section{Strangeon star merger tested by GW170817} \label{sec:3}

In the scenario that the pulsar-like compact stars could actually be strangeon stars,
the merging binary compact stars that triggers gravitational wave events as GW~170817
could then actually be binary strangeon stars.
In this section we will show the study on the parameter space of strangeon star model
according to the observation of GW170817 and possible future observations.

The most robust constraint that the binary strangeon star merger scenario has to confront,
is the tidal deformability constraint of GW170817. Mass quadrupole moment will be induced
by the external tidal filed of the companion during the late inspiral stage, accelerating the
coalescence, hence detectable by GW observations \cite{FlanaganHinderer}. This property of
the compact star can be characterized by the dimensionless tidal deformability
$\Lambda=(2/3)k_2/(GM/c^2R)^5$, where $k_2$ is the second tidal love number.

In order to study the parameter space of strangeon star model, we have calculated $k_2$
for a set of strangeon star EOSs with various choices of $U_0$ and $\rho_\mathrm{s}$. We have
followed the procedure as in \cite{Hinderer} to calculate $k_2$, namely,
introducing a static $l=2$ perturbation to the TOV equation and solving it with the strangeon star
EOSs. It's worth noting that due to finite surface density of strangeon star model,
a boundary treatment has to be done to ensure correct results \cite{Postnikov}.
In this study, we have explored parameter spaces with $U_0$ ranging from 20\,MeV to 100\,MeV
and $\rho_\mathrm{s}$ from 1.5 times to 2 times the nuclear density ($2.67\times10^{14}\,\mathrm{g/cm^3}$).
The TOV maximum mass with each EOS model is also calculated, as it's tightly related to the
post-merger evolution of the binary merger events.

Assuming both stars in the binary have low spins, the GW170817 observation translates into an
upper limit on the tidal deformability for a 1.4 solar mass star (labeled as $\Lambda(1.4)$)
of 800. Various studies on neutron star EOS models have been carried out based on this constraint, for example, a systematic study in \cite{Annala}. According to their results for neutron stars, the tidal deformability increases as the $M_\mathrm{TOV}$ increases. Consequently, the upper limit of $\Lambda(1.4)$
will rule out NS EOSs with $M_\mathrm{TOV}$ larger than 2.8 solar mass very robustly. According to our calculation in strangeon star model, the relationship between $\Lambda(1.4)$ and $M_\mathrm{TOV}$ still
holds qualitatively. However, the quantitative results change a lot. The largest possible $M_\mathrm{TOV}$
for the strangeon star EoSs preserving the $\Lambda(1.4)<800$ constraint is larger than 4\,$M_\odot$.
This quite large difference is resulted from the finite surface density of strangeon stars. Therefore, for
conventional quark star models which have a similar property, this quantitative difference is also found
in previous studies \cite{Zhou,Most}.

\begin{figure}
\resizebox{0.5\textwidth}{!}{%
  \includegraphics{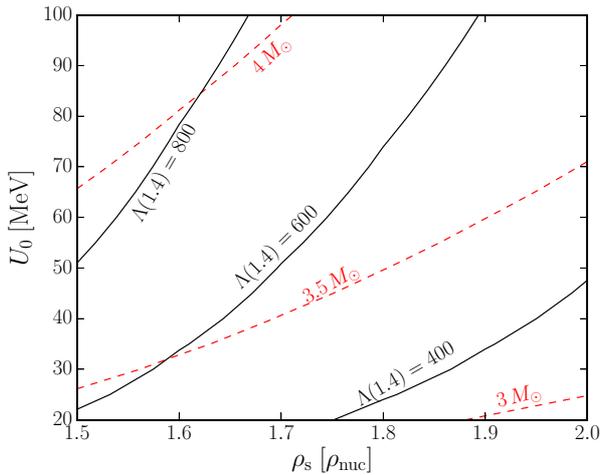}
}
\caption{Constraints on the equation of state parameters: $U_0$ and $\rho_\mathrm{s}$ (in unit of nuclear density with $\rho_\mathrm{nuc}=2.67\times10^{14}\,\mathrm{g/cm^3}$). Contours of the tidal deformability of a 1.4\,$M_\odot$ star ($\Lambda(1.4)$) are plotted in solid lines. According to the constraint of GW170817, any parameter choices below the top left solid contour is reasonable. Contours for the TOV maximum mass is also shown in dashed lines, although the strangeon star model is generally quite stiff. Hence the parameter choices will not be confronted by the observation of 2 solar mass pulsars \cite{Demorest,Antoniadis} in the parameter space we consider.}
\label{fig:1}       
\end{figure}

The details of our calculation result are shown in Fig.\ref{fig:1}.
The available parameter space is quite large as most of the EoS
models satisfy the tidal deformability constraint by GW170817. We
also show in the contour lines for $M_\mathrm{TOV}$ in
Fig.\ref{fig:1} to indicate the relation between $\Lambda(1.4)$ and
$M_\mathrm{TOV}$. As can be seen, both $M_\mathrm{TOV}$ and
$\Lambda(1.4)$ decrease as the surface density increase, which is
similar to the case of conventional quark stars described by MIT bag
model \cite{Zhou}. Whereas a larger $U_0$ makes the EoS stiffer,
resulting in a larger $M_\mathrm{TOV}$ and $\Lambda(1.4)$. For all
the models we have considered, the minimum $\Lambda(1.4)$ is
287\footnote{As a comparison, for NS models, $\Lambda(1.4)$ is 256
for the very soft EoS of APR4 (consists of $n,p,e,$ and
$\mu$\cite{APR4}), with $M_{\rm TOV} = 2.2M_\odot$.} with
$M_\mathrm{TOV}$ is 2.9\,$M_\odot$ (for the model with $U_0=20\,$MeV
and $\rho_\mathrm{s}=2\rho_\mathrm{nuc}$), which is still far beyond
the 2 solar mass constraint \cite{Demorest,Antoniadis}.
This sharp difference of $M_{\rm TOV}$ has clear consequence to the study of GRBs, as the post-merger should not be a black hole and would  power significantly both the GW170817-fireballs of GRB and kilonova in strangeon star model.

\section{Conclusions and discussions} \label{sec:4}

Bulk strong matter could be composed of strangeons, i.e.
quark-clusters with there-light-flavor symmetry of quarks, and
pulsar-like compact stars could actually be strangeon stars.
The EOS of strangeon stars is described in the Lennard-Jones model,
and the parameters $U_0$ and $\rho_{\rm s}$ are constrained by
both the tidal deformability $\Lambda$ of GW~170817 and $M_{\rm TOV}$.
We find that the available parameter space is quite large as most of
the EOS models satisfy the tidal deformability constraint by
GW170817.

Different from neutron stars, strangeon stars are self-bound rather
than gravity-bound. The finite surface density leads to a correction
to calculate the tidal deformability. As a result, they can reach a
much higher maximum mass under the same tidal deformability
constraint.
By contrast, it is not so easy for neutron star models to pass all
the tests. For example, according to~\cite{Annala}, neutron stars
cannot reach higher than $2.8 M_\odot$ in order to satisfy the
constraint of tidal deformability.

The parameters $U_0$ and $\rho_{\rm s}$, which characterize the
inter-strangeon potential and determine the EOS of strangeon stars,
should have implications on the properties of strong interaction at
the low energy scales.
From the constraints by both GWs ($\Lambda \leq 800$) and the mass
measurement ($M_{\rm TOV}\geq 2 M_\odot$), the allowed region of parameters is still very large.
We may expect $U_0<60$ MeV and $\rho_{\rm s}>1.5$ times of nuclear
density since the detected masses of stellar black holes are usually
lager than $4M_\odot$ at least~\cite{Bailyn1998,Farr2011}.
Future GW detections for smaller values of $\Lambda$ along with larger values of
$M_{\rm TOV}$ will be helpful to make better constraints on the strangeon star model.

All EOSs we choose here lead to values of $M_{\rm TOV}$ far beyond $2 M_\odot$,
indicating that all of the known pulsar-like compact stars are far below the maximum mass.
High maximum mass also indicates quite a different scenario for the post-merger phase.
A much longer lived strangeon star as the merger remnant should be expected.
This long-live remnant could be helpful to understand the GW~170817 associated
kilonova observation AT 2017gfo~\cite{Lai2018RAA,Yu2017,Li2018}. The continuous
energy injection from the spin down power of the merger remnant is a natural energy source
for the extended emission of AT2017\,gfo, without requiring larger opacity and larger amount of
ejecta mass compared with numerical simulation of binary mergers. Particularly, it is
hinted that there might be an X-ray flare related to the central engine after more than 100 days
of the merger \cite{Zhang}, which highly favors the possibility that the remnant has not collapsed
to a black hole yet. The strangeon star model will allow for such a long lifetime for the merger
remnant even for the model with the smallest $M_\mathrm{TOV}$.

Additionally, as mentioned above, isolate strangeon star, or binary
strangeon stars in the early inspiral stage when they are separated
far enough, could be in solid state, for which the tidal
deformability could be much smaller or even negligible than the
values estimated with perfect fluid energy momentum tensor.
Depending on the breaking strain ($\sigma$) and shear modulus
($\mu$) of the solid structure, the tidal heating effect might melt
the solid star at a certain breaking frequency \cite{Postnikov}.
\begin{eqnarray}
f_{\mathrm{br}}&=&(\frac{2}{3})^{1/4}\frac{1}{\pi}(\frac{Q_{22\mathrm{max}}}{\lambda})\nonumber
\\
&=&20\times(\frac{Q_{22\mathrm{max}}}{10^{40}\mathrm{\,g\,cm}^2})^{1/2}
(\frac{\lambda}{2\times
10^{36}\mathrm{g\,cm}^2\mathrm{s}^2})^{-1/2}\,\mathrm{Hz}
\end{eqnarray}
in which $\lambda$ is the tidal deformability resuming the
dimensional units and $Q_{22\mathrm{max}}$ is the maximum quadrupole
moment that should be induced in the solid star before it is melt,
which can be estimated as \cite{Owen}
\begin{eqnarray}
&&Q_{22\mathrm{max}}=2.8\times 10^{41} \nonumber \\
&&\frac{\mu}{4\times 10^{32}\mathrm{erg\,cm}^{-3}}
(\frac{R}{10\,\mathrm{km}})^6(\frac{M}{1.4\,M_{\odot}})^{-1}\frac{\sigma_{\mathrm{max}}}{0.01}\,\mathrm{g\,cm}^2
.
\end{eqnarray}
As a result, if indeed isolated strangeon stars are in solid state,
we might be able to observe a sudden change in the tidal
deformability at a certain gravitational wave frequency in future
observations. The breaking frequency itself will also provide
important information about the properties of the solid star. This
should be studied in more details in future work.

The state of supranuclear matter in compact stars essentially
relates to the fundamental strong interaction at the low energy
scales, which still remains a challenge.
The strangeon star model perceives a pulsar-like compact star as a gigantic strange nucleus whose building blocks are strangeons.
Up to now, the strangeon star model has passed all of the
observational tests, and we expect that the more advanced GW
observations in the future would tell us more about the strangeon
stars and the bulk strong matter.

\acknowledgement{This work is supported by the National Key R\&D
Program of China (No. 2017YFA0402602), the National Natural Science
Foundation of China (Grant Nos. 11673002 and U1531243), and the
Strategic Priority Research Program of CAS (No. XDB23010200). Xiaoyu
Lai is supported by the Science and technology research project of
Hubei Provincial Department of Education (No. D20183002).}

%
%

\end{document}